\documentclass[journal]{IEEEtran}
\ifCLASSINFOpdf
  \usepackage[pdftex]{graphicx}
\else
\fi
\usepackage[cmex10]{amsmath}
\usepackage{algorithmic}

\usepackage{array}

\usepackage[utf8]{inputenc}

\usepackage[tight,footnotesize]{subfigure}
\begin{document}
%
\title{Reaching Unanimous Agreements within Agent-Based Negotiation Teams with Linear and Monotonic Utility Functions}
%
%
%

\author{Victor~Sanchez-Anguix,
        Vicente~Julian,
        Vicente~Botti,
        and~Ana~García-Fornes
\thanks{V. Sanchez-Anguix, V. Julian, V. Botti, and A. Garcia-Fornes are with the Departamento de Sistemas Informáticos y Computación, Universidad Politécnica de Valencia, Valencia,
Spain, Camí de Vera s/n, 46022 e-mails: \{sanguix,vinglada,vbotti,agarcia\}@dsic.upv.es}
\thanks{This work is supported by TIN2008-04446, PROMETEO/2008/051, TIN2009-13839-C03-01, CSD2007-00022 of the Spanish government, and FPU grant AP2008-00600 awarded to Víctor Sánchez-Anguix.}%
}

\maketitle

\begin{abstract}
In this article, an agent-based negotiation model for negotiation teams that negotiate a deal with an opponent is presented. Agent-based negotiation teams are groups of agents that join together as a single negotiation party because they share an interest that is related to the negotiation process. The model relies on a trusted mediator that coordinates and helps team members in the decisions that they have to take during the negotiation process: which offer is sent to the opponent, and whether or not the offers received from the opponent are accepted. The main strength of the proposed negotiation model is the fact that it guarantees unanimity within team decisions since decisions report a utility to team members that is greater than or equal to their aspiration levels at each negotiation round. This work analyzes how unanimous decisions are taken within the team and the robustness of the model against different types of manipulations. An empirical evaluation is also performed to study the impact of the different parameters of the model. 
\end{abstract}


%
\IEEEpeerreviewmaketitle

\section{Introduction}
%
%
%
%
\IEEEPARstart{A}{} negotiation team is a group of two or more interdependent individuals who join together as a single negotiation party because of their similar interests and objectives related to the negotiation and who are all present at the bargaining table \cite{thompson01}. Therefore, this group of individuals unites because their members share goals that are related to a negotiation with an opponent. For instance, negotiation teams formed by different stakeholders are usually sent to the negotiation table  when a company decides to sell a product line to another company. Nevertheless, as it has been stated in social sciences, negotiation teams are not necessarily unitary players since team members may have different preferences regarding the possible outcomes of the negotiation process \cite{halevy08}. Thus, given the divergence in preferences between teammates, the team has to agree upon, not only a negotiation strategy to carry out with the opponent, but also upon those agreements that are acceptable to the team. Despite being studied in social sciences to some extent \cite{thompson01,halevy08}, as far as we know, negotiation teams have been overlooked by artificial intelligence research. We argue that a negotiation team is also an element that may be appropriate for some scenarios involving software agents. For instance, let us imagine an example based on an electronic market for travel and tourism. In this system, a group of friends (each friend is represented by a software agent) has decided to go on a trip together. This goal requires a negotiation with a travel agency agent. The fact that the group of agents (group of friends) has a common and shared goal (which is going on a trip together), is clear, and it requires an agreement with an opponent (the travel agency agent). However, it also seems reasonable to assume that friends may have different preferences regarding the negotiable trip conditions: hotel quality, price, number of days to stay, etc. For example, while some friends may care more about comfort, others may be more interested in money. In this type of scenario, and specially in open multi-agent systems, mediated preference aggregation is complicated since i) agents may be inclined to exaggerate their preferences in order to ensure a certain level of utility; ii) preferences are delicate information which may not be revealed to anyone; iii) utility functions may be different and require an extensive and costly aggregation. Hence, mechanisms that allow an agent-based negotiation team to handle intra-team conflict (divergences in preferences) while trying to get a deal from travel agencies' agents are needed. Thus, it is necessary to provide agent-based models for negotiation teams.

This article describes a mediated negotiation model for agent-based negotiation teams that negotiate with an opponent. The negotiation model defines the communications protocol with the opponent, what decisions are taken by the negotiation team, and how and when these decisions are taken  (i.e., team dynamics) \cite{sanchez-anguix10}. More specifically, our preliminary study presented about this model (Full Unanimity Mediated or FUM) in \cite{sanchez-anguix11} is extended further. FUM is able to guarantee unanimity in decisions taken within the negotiation team as long as team members share the same type of monotonicity for valuation functions of the attributes. This assumption is relatively natural in buyer-seller settings found in electronic commerce (e.g., a team of buyers may value with the same type of valuation function attributes like the price, the quality of the product, and the time of dispatch). The proposed model relies on unanimity rules regarding the opponent's offer acceptance and an iterated offer construction process to determine which offer is sent to the opponent. This article is organized as follows. First, we describe our negotiation model. Then, we analyze how unanimity is assured within the team as well as the robustness of this model against different types of attacks. Then, we evaluate the empirical response of the proposed model depending on the impact of the different model parameters and we analyze possible incentives that team members may have to deviate from the proposed behavior. We then relate our work to other works found in artificial intelligence. Finally, we summarize the conclusions of this work and discuss our future research.  

\section{Negotiation Model}
\label{sec:model}
Traditionally, a negotiation model is composed of a negotiation protocol, which defines the set of actions that are available for agents at each instant, and the negotiation strategy, which defines the decision making mechanisms employed by agents during the negotiation. In this article, we present a negotiation model where a negotiation team negotiates with an opponent. Despite resembling a bilateral negotiation scenario, negotiations that have teams as participants are slightly different since team dynamics also play a key role. Thus, three different elements have to be specified when a team negotiation model is proposed: negotiation protocol with the opponent, the negotiation strategy used by the opponent, and the intra-team negotiation strategy followed by team members in order to decide the actions to perform during the negotiation process. An intra-team strategy defines which decisions are taken by the team, and how and when these decisions are taken. More specifically, it is defined by the negotiation protocol followed among team members within the team and by the strategies followed by agents within the team. In this section, the general assumptions of our negotiation model and our negotiation model itself are described. Special attention is focused on the interactions among team members, which are carried out before and during the negotiation process and the negotiation strategy followed by team members within the team. 

\subsection{General Assumptions}
\begin{itemize}
 
 \item  In our model, a group of agents has formed a team $A=\{a_{1},a_{2},...,a_{M}\}$ whose goal is to negotiate a successful deal with an opponent $op$. However, each team member $a_{i}$ may have different preferences about the negotiation issues.
 \item Communications between the team and the opponent are carried out by means of a mediator that is trusted by the team. This mediator sends team decisions to the opponent and receives, and later broadcasts, decisions from the opponent to team members. Thus, the fact that the opponent is communicating with a team is not known by the opponent, which only interacts with the trusted mediator. The trusted mediator also performs other tasks that allow team members to reach unanimous decisions regarding the offer that is sent to the opponent and whether or not the opponent's offer is accepted.
 \item The negotiation domain is comprised of $n$ real-valued attributes whose domain is $[0,1]$. Thus, the possible number of offers is $[0,1]^{n}$.
\item A complete offer is represented as $X=\{x_{1},x_{2},...,x_{n}\}$, where $x_{i}$ is the value assigned to the i-th attribute. The notation $X_{i\rightarrow j}^{t}$ is employed to indicate that $X$ is the offer sent by agent/team $i$ to agent $j$ at round $t$.
 \item Team composition will remain static during the negotiation process. It is acknowledged that team members may leave or join the group in certain specific situations. However, membership dynamics is not considered in this article, and it is designated as future work.
 \item All of the agents use linear utility functions to represent their private preferences. Negotiation attributes are supposed to be independent. Thus, the value of a specific attribute does not affect the valuation of other attributes' values. These functions can be formalized as follows:

 \begin{equation}
 U_{i}(X)= \overset{n}{\underset{j=1}{\sum}} w_{i,j}\;V_{i,j}(x_{j})
 \end{equation}
 where $V_{i,j}(.)$ is a monotonic valuation function that transforms the attribute value to $[0,1]$, and $w_{i,j}$ is the weight or importance that is given by the agent $i$ to the j-th attribute. Weights are normalized so that $\sum_{j=1}^{n} w_{i,j}=1 $ holds for every utility function. It is assumed that teammates share the same type of monotonic valuation function (either increasing or decreasing) for each negotiation attribute. In contrast to team members, the opponent valuation function is always the opposite type of monotonic function. Thus, if team members employ monotonically increasing functions, the opponent will be modeled using monotonically decreasing functions. It is reasonable to assume this model for valuation functions in e-commerce scenarios. Buyers usually share the same type of valuation function for attributes such as the price (monotonically decreasing), product quality (monotonically increasing), and the dispatch time (monotonically decreasing), whereas sellers usually use the opposite type of monotonic functions (monotonically increasing for price, monotonically decreasing for product quality, and monotonically increasing for dispatch time). As for attributes' weights, it is considered that each team member may assign different weight/importance to each negotiation issue. Therefore, differences among teammates are introduced by assigning different weights to negotiation attributes. Nevertheless, it should be highlighted that since team members share the same type of monotonic function, if one of the team members increases its welfare by increasing/decreasing one of the attribute values, the other team members will stay at the same welfare level or they will also increase their welfare. Thus, there is potential for cooperation among team members. Weights given by the opponent to attributes may also be different to those given by teammates. Agents do not know the form of other agents' utility functions, even if they are teammates. 
 
 \item The opponent has a private deadline $T_{op}$, which defines his maximum number of negotiation rounds. Once $T_{op}$ has been reached in the negotiation process, the opponent will exit the process and the negotiation will end in failure. The team has a private joint deadline $T_{A}$, which is common information for team members. Once this deadline has been reached, the team will exit the negotiation process and the negotiation will end in failure. We consider that $T_{A}$ has been agreed upon by team members before the negotiation process starts.

\item The opponent has a reservation utility $RU_{op}$. Any offer whose utility is lower than $RU_{op}$ is rejected. Each team member has a private reservation utility $RU_{a_{i}}$, where $a_{i}$ is a team member. This individual reservation utility is not shared among teammates. Therefore, a team member $a_{i}$ rejects any offer whose value is under $RU_{a_{i}}$. In this setting, reservation utilities represent the individual utility of each agent if the negotiation process fails. 

\end{itemize}

\subsection{Negotiation Protocol with the Opponent}
In this section, the negotiation process between a negotiation team and an opponent is studied. The fact that one of the parties is a team is transparent to the other party. Thus, superficially, the scenario resembles a bilateral negotiation scenario. Because of this, we decided to model the interaction between the team and the opponent as an alternating bilateral negotiation process \cite{rubinstein82}. In this protocol, one of the two agents is the initiator and sends the first offer to the other party or responding agent. The responding agent receives the offer and decides whether he/she accepts the offer or he/she sends a counter-offer as response. If the responding agent sends a counter-offer, the initiator agent has to decide whether he/she accepts the counter-offer or not. If the counter-offer is rejected, a new round starts and the process is repeated again until a deal is accepted (successful negotiation) or one of the parties decides to quit the negotiation since its deadline has been reached (failed negotiation). In our negotiation model, we consider that a trusted mediator is responsible for sending team decisions to the opponent and broadcasting opponent decisions to team members. Nevertheless, this mediator not only acts as a coordinator but also helps the team to reach unanimous agreements by means of an iterated process for offer generation and unanimity rules for opponent offer acceptance.  

\subsection{Opponent Negotiation Strategy}
A negotiation strategy defines the decision-making of an agent in a negotiation process. In this case, the negotiation strategy is constituted by the concession strategy, which marks the aspiration level of the opponent in terms of utility at each negotiation round, and the acceptance criterion, which determines whether the team offer is accepted or not. 
\begin{itemize}
 \item It is assumed that the opponent uses a concession strategy to carry out during the negotiation process. A concession strategy typically (although not necessarily) starts by demanding the maximum aspiration and, as the negotiation process advances, the aspiration demanded tends to be lowered. The amount of concession/reduction applied at each step may depend on the specific tactic selected by the opponent. In this article, our main focus of interest is the general behavior of the proposed intra-team strategy. Thus, a set of well-known negotiation tactics was selected as the opponent negotiation strategy: time-dependent tactics \cite{faratin98,lai08}. We formalized time-dependent tactics as suggested by \cite{lai08}:
 \begin{equation}
 \label{time}
  s_{op}(t)=1-(1-RU_{op})(\frac{t}{T_{op}})^{\frac{1}{\beta_{op}}}
 \end{equation}
 where $t$ is the current negotiation round and $\beta_{op}$ is a negotiation strategy parameter (concession speed) which determines how  concessions are made towards $RU_{op}$. On the one hand, when $\beta_{op}=1$ the concession is linear and each negotiation round the same amount of concession is performed, and when $\beta_{op}<1$ the concession is boulware and very little is conceded at the start of the negotiation process but the agent concedes faster as the negotiation deadline approaches. On the other hand, when $\beta_{op}>1$ the concession is conceder and the agents concede fast towards the reservation utility in the first rounds. 
 \item The opponent uses an offer acceptance criterion $ac_{op}(.)$ during the negotiation process. It is formalized as follows:
 \begin{equation}
ac_{op}(X) = \left\lbrace
	  \begin{array}{l l}
	     accept & \mbox{if } s_{op}(t+1) \leq U_{op}(X) \\
	     reject & \mbox{otherwise}
	  \end{array}
	  \right.
 \end{equation}
 where $t$ is the current round, $X$ is the offer received from the team, $U_{op}(.)$ is the utility function of the opponent, and $s_{op}(.)$ is the opponent concession strategy. Thus, an offer is accepted by $op$ if it reports a utility that is equal to or greater than the utility of the offer that $op$ would propose in the next round.
\end{itemize}

\subsection{Intra-Team Strategy: Negotiation Protocol within the Team}
\begin{figure}
\begin{center}
\includegraphics[width=0.93\linewidth]{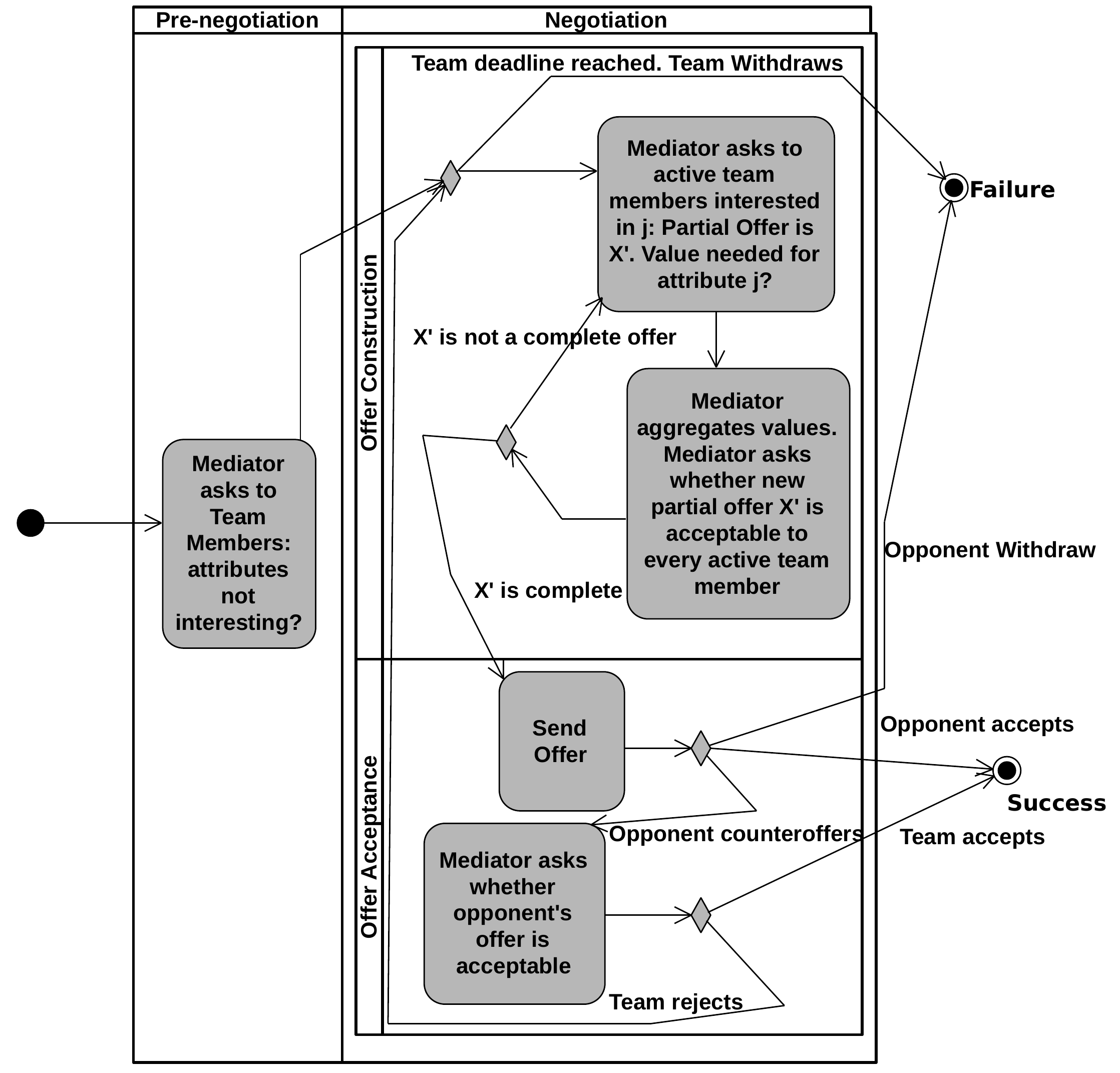}
\caption{This figure represents an activity diagram of the general process followed by the team in our model. }
\label{diagram}
\end{center}
\end{figure}

The negotiation protocol followed by team members for team communications can be divided into two different phases: the protocol followed during the pre-negotiation, and the protocol used during the negotiation process. Both of them are thoroughly described below. A general overview of the process followed by team members can be observed in Fig. \ref{diagram}.

\subsubsection{Pre-negotiation}
In the pre-negotiation, team members confidentially share certain information about their preferences with the trusted mediator. Each team member specifies which attributes' decision rights it is willing to hand over when the team proposes an offer. It is reasonable that each team member may be willing to sacrifice decision rights pertaining to negotiation attributes that have little importance or no importance at all for one's own interests. This decision may help to find a more satisfactory agreement for the opponent while maintaining good quality for one's own utility. The fact that some attributes may yield little or no importance at all for some team members is also feasible in a team setting, since some of these attributes may have been introduced to satisfy the interests of a subgroup of team members. 

Therefore, for each team member $a_{i}$, the trusted mediator asks the set of attributes $NI_{a_{i}}$ whose decision rights conform the set of rights that $a_{i}$ is willing to hand over when building an offer for the opponent. This information is annotated by the mediator in an interest matrix $I$. Each matrix position $I(i,j)$ indicates whether the team member $a_{i}$ holds decision rights for attribute $j$ ($I(i,j)=true$) or not ($I(i,j)=false$). How each team member $a_{i}$ specifies this set of attributes $NI_{a_{i}}$ is described later in Subsection \ref{subsec:intra-strategy}. The communication protocol carried out during the pre-negotiation phase,  described from the point of view of the mediator, is shown in Fig. \ref{op_pre-negotiation}.

\begin{figure}
\begin{algorithmic}
\STATE \COMMENT{Start pre-negotiation phase}
\STATE $\forall i,j$, $I(i,j)=true$
\STATE Ask for $NI_{a_{i}}$ to each $a_{i}$
\STATE Receive responses $NI_{a_{i}}$ from each $a_{i}$
\FOR{ $a_{i} \in A$}
\FOR{ $j \in NI_{a_{i}}$}
 \STATE $I(i,j)=false$
\ENDFOR
\ENDFOR
\STATE \COMMENT{End pre-negotiation phase}
\end{algorithmic}
\caption{Pre-negotiation protocol followed by team $A$ and the trusted mediator. In this schema, we show the protocol from the point of view of the mediator. The protocol followed by team members is analogous and straight-forward.}
\label{op_pre-negotiation}
\end{figure}

\subsubsection{Negotiation}
\begin{figure}[t]
\begin{center}
\includegraphics[width=\linewidth]{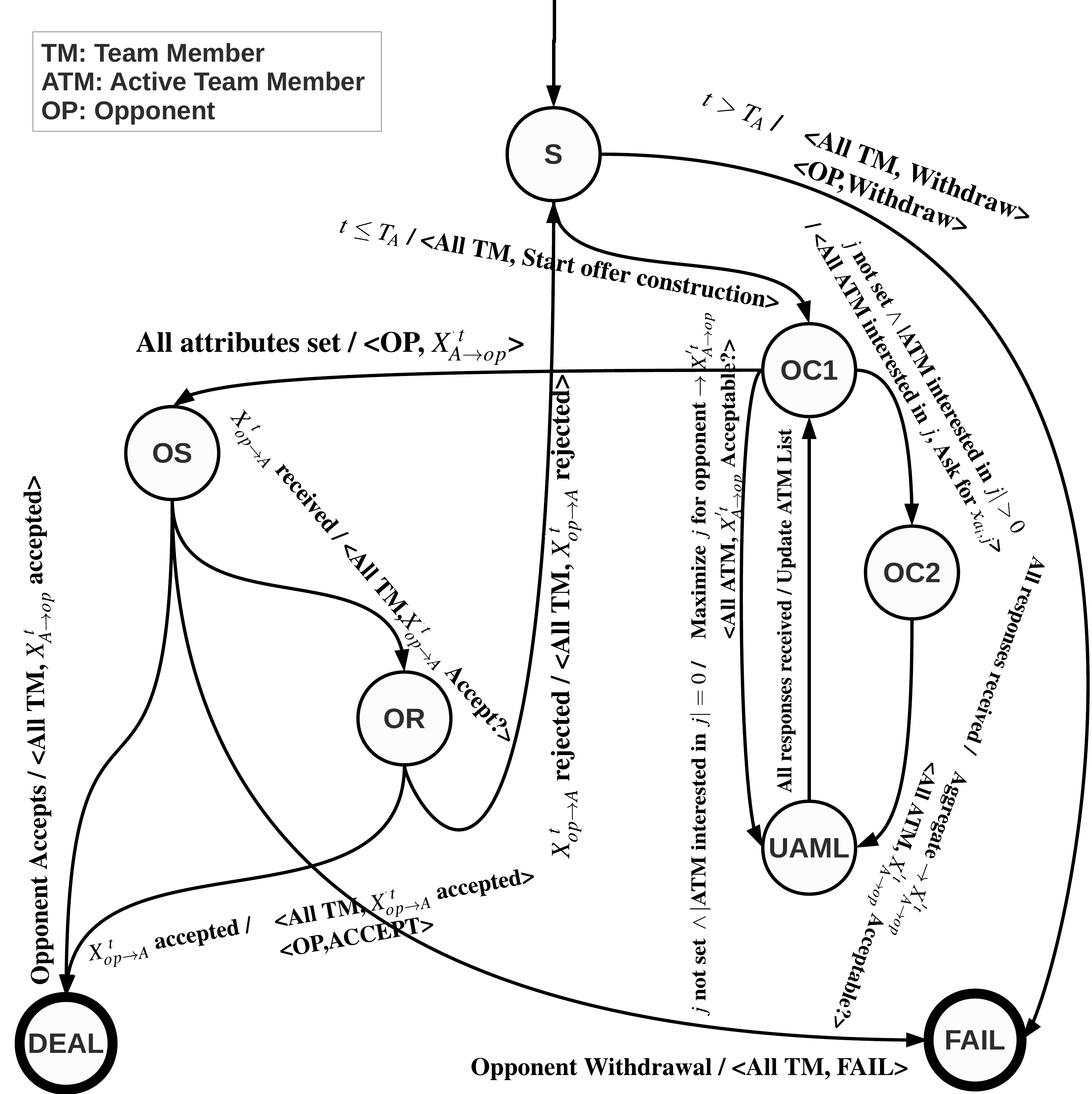}
 \caption{Finite state machine formalization of the negotiation from the point of view of the mediator. The transitions follow the format \textit{Events / Actions}. Messages sent by the mediator follow the format \textit{$<$Recipient,Message$>$}.}
\label{fsm}
\end{center}
\end{figure}

Three possible actions can be carried out by the negotiation team at each negotiation round $t$: (a) accept/reject opponent offer $ac_{A}(X_{op\rightarrow A}^{t})$ (Offer acceptance in Fig. \ref{diagram}); (b) send an offer/counter-offer $X_{A\rightarrow op}^{t}$ (Offer construction in Fig. \ref{diagram}); (c) abandon the negotiation process. This last action is performed when the team deadline $T_{A}$ is reached. Thus, when $t>T_{A}$, the mediator informs the opponent about the team's withdrawal. The mediator also has a very important role in the coordination mechanisms employed by the team to decide upon action (a) and (b). The coordination processes (a) and (b) are described in a detailed way. Furthermore, a finite state machine formalization of the negotiation from the point of view of the mediator can be observed in Fig. \ref{fsm}. It shows the offer construction and the opponent offer acceptance processes.

\paragraph{Offer acceptance}
When team A has to decide whether or not to accept the opponent's offer ($ac_{A}(X_{op\rightarrow A}^{t})$), first, the mediator receives the opponent's offer $X_{op\rightarrow A}^{t}$. This offer is publicly announced to all of the team members by the mediator. Then, the mediator opens a private voting process where each team member $a_{i}$ should specify whether or not it supports acceptance of the opponent's offer $ac_{a_{i}}(X_{op\rightarrow A}^{t})$. How each team member decides whether or not the opponent offer is supported will be described in Subsection \ref{subsec:intra-strategy}. Once every vote has been received, the mediator counts the number of positive votes (votes that support the opponent's offer). The offer is accepted if the number of positive votes is equal to the number of team members. Otherwise, the offer is rejected. A complete view of this communication protocol can be observed in Fig. \ref{op_neg_ac}.
\begin{figure}
\begin{algorithmic}
\STATE Receive $X_{op\rightarrow A}^{t}$
\STATE Broadcast $X_{op\rightarrow A}^{t}$ in $A$ (start voting process)
\STATE Receive response $ac_{a_{i}}(X_{op\rightarrow A}^{t})$ from each team member $a_{i}$ in $A$
\STATE $votes=0$
\FOR{ $a_{i} \in A$}
 \IF{$ac_{a_{i}}(X_{op\rightarrow A}^{t})=true$}
  \STATE $votes=votes+1$
 \ENDIF
\ENDFOR
\IF{$votes=|A|$}
 \STATE $ac_{A}(X_{op\rightarrow A}^{t})=true$
 \ELSE
  \STATE $ac_{A}(X_{op\rightarrow A}^{t})=false$
 \ENDIF
\end{algorithmic}
\caption{Negotiation protocol followed by team $A$ and the trusted mediator to decide the acceptance of the opponent offer. In this schema, we show the protocol from the point of view of the mediator. The protocol followed by team members is analogous and straight-forward.}
\label{op_neg_ac}
\end{figure}

\paragraph{Offer construction}
The mediator coordinates an iterated offer-building process in order to ensure unanimity in the offer $X_{A\rightarrow op}^{t}$ sent to the opponent. For that purpose, each attribute value is adjusted one by one. Before the iterated process starts, the mediator considers every team member $a_{i}$ as an active member in the offer construction process. Once the iterated offer construction process starts, the trusted mediator selects an attribute $j$ from the set of attributes that have not yet been set.  Given the partial offer $X_{A\rightarrow op}^{'t}$ built until the moment, the mediator asks each active team member $a_{i}$ who is interested in $j$ ($I(i,j)=true$) about the value $x_{a_{i},j}$ needed to get as close as possible to its current aspiration level $s_{a_{i}}(t)$. When private responses have been gathered from every team member, the mediator decides a value $x_{j}$ for the attribute $j$. Here, the morphology of the proposed utility functions comes into play. Due to the fact that team members share the same type of monotonicity for valuation functions, the trusted mediator can aggregate agents' opinions by means of the $max$ function (monotonically increasing) or the $min$ function (monotonically decreasing). As will be proved in Section \ref{sec:theo}, the value decided for $x_{j}$ ensures unanimity among team members under certain assumptions. The value $x_{j}$ is set in a new partial offer $X_{A\rightarrow op}^{'t}$ which is publicly announced to team members. Then, the mediator asks every active agent in the offer construction process whether or not the new partial offer is satisfactory at round $t$, $ac'_{a_{i}}(X_{A\rightarrow op}^{'t})$. Those agents that agree with the current state of $X_{A\rightarrow op}^{'t}$ are eliminated from the active list. Those attributes which are not interesting for any team member are maximized according to the opponent's preferences. The process steps back to the selection of a new attribute $j$ until all of them have been set. A more detailed description of this process can be observed in Fig. \ref{op_neg_partial}. As will be reviewed in Section  \ref{sec:theo}, when agents comply with certain assumptions, the proposed iterated process is able to reach an offer that is supported by all of the team members at each negotiation round (unanimity). Obviously, the resultant offer depends on the agenda of attributes employed by the mediator. Since unanimity is guaranteed independently of the agenda employed by the mediator, the offer constructed should be as satisfactory as possible for the opponent. Ideally, the team should try to fulfill its own interests with those attributes that are less important for the opponent.

\begin{figure}
\begin{algorithmic}
\STATE $X_{A\rightarrow op}^{'t}=\emptyset$
\STATE $A'=A$
\STATE \COMMENT{For each attribute}
\FOR{$j \in N$}
\STATE $V=\emptyset$
\STATE \COMMENT{Check opinion of team members who are active in the building phase and are interested in the attribute}
\FOR{$a_{i} \in A' \wedge I(i,j)=true$}
\STATE Ask for $x_{a_{i},j}$
\STATE Receive $x_{a_{i},j}$
\STATE $V=V\bigcup x_{a_{i},j}$
\ENDFOR
\STATE \COMMENT{Aggregate agents' opinions}
\IF{$|V| = 0$}
\STATE $x_{j}=$ best\_value\_for\_opponent(j)
\ELSE
\STATE $x_{j}=max(V)\;\;or\;\;min(V)$
\ENDIF
\STATE $X_{A\rightarrow op}^{'t}=X_{A\rightarrow op}^{'t}\bigcup x_{j}$
\STATE Make public new $X_{A\rightarrow op}^{'t}$ among team members
\STATE \COMMENT{Update list of agents who are active in the building phase}
\FOR{$a_{i} \in A'$}
\STATE Ask for $ac'_{a_{i}}(X_{A\rightarrow op}^{'t})$
\STATE Receive $ac'_{a_{i}}(X_{A\rightarrow op}^{'t})$
\IF{$ac'_{a_{i}}(X_{A\rightarrow op}^{'t})=true$}
 \STATE $A'=A'-a_{i}$
\ENDIF
\ENDFOR
\ENDFOR
\STATE $X_{A\rightarrow op}^{t}=X_{A\rightarrow op}^{'t}$
\end{algorithmic}
\caption{Negotiation protocol followed by team $A$ and the trusted mediator to build an offer to be sent to the opponent. In this schema, we show the protocol from the point of view of the mediator. The protocol followed by team members is analogous and straight-forward.}
\label{op_neg_partial}
\end{figure}

\subsection{Intra-Team Strategy: Team Members' Strategy within the Team}
\label{subsec:intra-strategy}
This subsection specifies how team members answer the mediator's petitions. On the one hand, during the pre-negotiation, team members decide upon which attribute decision rights are handed over. On the other hand, during the negotiation, agents have to decide whether they accept the opponent's offer and which values should be set for the offer to be sent to the opponent. The behavior of team members in these decision making processes is described below.

\subsubsection{Pre-negotiation}
In the pre-negotiation phase, the mediator asks each team member the set of attributes' decision rights that it is willing to hand over. Its size may range from 0 attributes to the whole set of attributes. How many decision rights $a_{i}$ is willing to hand over depends on an individual and private value $\epsilon_{a_{i}} \in [0,1]$. When $\epsilon_{a_{i}}=0$, the agent is only willing to hand over rights that yield no interest at all (i.e., attributes $j$ whose $w_{a_{i},j}=0$), whereas when $\epsilon_{a_{i}}=1$, the agent is willing to hand over all of the attributes' decision rights. The set of attributes $NI_{a_{i}}$ whose decision rights are handed over by agent $a_{i}$ follow this equation: 
\begin{equation}
\label{eq_team_tolerance}
 \underset{j \in NI_{a_{i}}}{\sum}w_{a_{i},j} \leq \epsilon_{a_{i}}
\end{equation}

Thus, $\epsilon_{a_{i}}$ acts as an upper limit that determines the total importance given by $a_{i}$ to the attributes whose decision rights are handed over. Given a value for $\epsilon_{a_{i}}$, there are multiple sets $NI_{a_{i}}$ that fulfill Eq. \ref{eq_team_tolerance}. A reasonable heuristic is to assume that the agent is willing to concede as many decision rights as possible since this will enhance the possibility of finding an agreement with the opponent. Hence, each team member $a_{i}$ chooses the largest possible set $NI_{a_{i}}$ that fulfills Eq. \ref{eq_team_tolerance}.
\subsubsection{Negotiation}
In the negotiation process, two different decisions are taken by the team: whether or not they accept the opponent's offer, and which offer is sent to the opponent.

First, the decision making mechanism $ac_{a_{i}}(.)$, used by each agent $a_{i}$ to decide whether or not it supports the opponent's offer, is described. It seems appropriate to assume that the agent will accept the opponent's offer if it reports a utility that is greater than or equal to the aspiration level marked by the concession strategy in the next round:
\begin{equation}
\label{op_ac}
ac_{a_{i}}(X) = \left\lbrace
	  \begin{array}{l l}
	     true & \mbox{if } s_{a_{i}}(t+1) \leq U_{a_{i}}(X) \\
	     false & \mbox{otherwise}
	  \end{array}
	  \right.
\end{equation}
where $true$ means that the agent supports the opponent's offer, $false$ has the opposite meaning, and $s_{a_{i}}(.)$ is the concession strategy employed by agent $a_{i}$ to calculate the aspiration level at each negotiation round $t$. Regarding the concession strategy employed by team members, it is considered that team members have agreed upon a time-based concession strategy with a common $\beta_{A}$. Thus, the concession strategy $s_{a_{i}}(.)$ followed by each team member $a_{i}$ can be formalized as depicted below. It is a modification of the well-known concession strategy used in Equation \ref{time}.
\begin{equation}
 s_{a_{i}}(t)=(1-\epsilon_{a_{i}})-(1-\epsilon_{a_{i}}-RU_{a_{i}})(\frac{t}{T_{A}})^{\frac{1}{\beta_{A}}}
\end{equation}
For the expression above, it can be observed that each agent's aspiration level, despite being governed by the same $\beta_{A}$, depends on the private reservation utility of each agent $RU_{a_{i}}$. $\epsilon_{a_{i}}$ acts as a limit for the maximum utility demanded by the concession strategy. Since the agent has handed over decision rights for a set of attributes whose weights sum up to $\epsilon_{a_{i}}$, the maximum utility that the agent is able to demand by itself is $(1-\epsilon_{a_{i}})$. This is reflected in the equation above.

\begin{figure*}
\centering
 \includegraphics[width=0.95\linewidth]{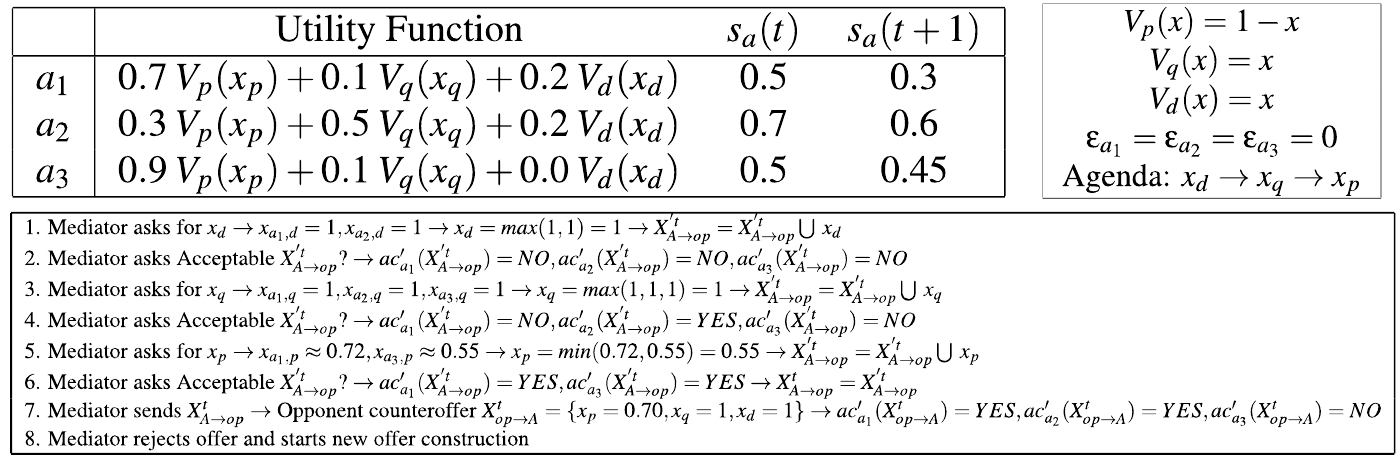}
 \caption{The figure shows a simplistic trace of the proposed model during one negotiation round $t$. The team is composed of three buyers $(a_{1},a_{2},a_{3})$, whereas the opponent is a seller. The upper left table shows the utility functions of team members and their aspiration levels at round $t$ and $t+1$. The upper right table shows team members' valuation functions for negotiation attributes and the agenda learnt by the mediator. The negotiation attribute are $p$ (price, monotonically decreasing), $q$ (quality, monotonically increasing), and $d$ (payment date time, monotonically increasing). The lower table is the trace of the negotiation round. In (1), the first attribute of the agenda $x_{d}$ is set, but it should be highlighted that $a_{3}$ does not participate in its construction since $\epsilon_{a_{3}}=0$ and, thus, $a_{3}$ has handed the decision rights over $x_{d}$ during the pre-negotiation. After (4), $a_{2}$ does not participate in the offer construction since it has reached its desired aspiration with the value set for attributes $x_{d}$ and $x_{q}$. Finally, in (7) the opponent's counteroffer is rejected since the offer is not acceptable for $a_{3}$.}
 \label{fig:trace}
\end{figure*}

Second, in the case of the iterated construction process, team members take two decisions: which value $x_{a_{i},j}$ is requested for attribute $j$ given the partial offer $X_{A\rightarrow op}^{'t}$, and whether or not the new partial offer is acceptable $ac'_{a_{i}}(X_{A\rightarrow op}^{'t})$. When requesting a value for $j$, each team member communicates anonymously the value $x_{a_{i},j}$ which gets as close as possible to its desired aspiration level $s_{a_{i}}(t)$. This value can be calculated by obtaining the attribute value $x_{a_{i},j}$ whose weighted utility ($w_{a_{i},j}V_{a_{i},j}(x_{a_{i},j})$) is the closest to the utility needed by the partial offer in order to reach the desired utility level ($s_{a_{i}}(t)-U_{a_{i}}(X_{A\rightarrow op}^{'t})$) :
\begin{equation}
\label{eq:bid1}
 x_{a_{i},j}= \underset{x\in [0,1]}{\mbox{argmin }} (s_{a_{i}}(t)-U_{a_{i}}(X_{A\rightarrow op}^{'t})-w_{a_{i},j}V_{a_{i},j}(x))
\end{equation}
where $x_{a_{i},j}$ is set so that the new offer's utility does not exceed the aspiration level marked by the concession strategy:
\begin{equation}
\label{eq:bid2}
 s_{a_{i}}(t)-U_{a_{i}}(X_{A\rightarrow op}^{'t})-w_{a_{i},j}V_{a_{i},j}(x_{a_{i},j})\geq 0
\end{equation}
Once the partial offer $X_{A\rightarrow op}^{'t}$ has been updated by the mediator, those agents that are still active in the construction process are asked whether or not the new offer is acceptable for the current negotiation round. Again, we consider that a partial offer is acceptable for an agent $a_{i}$ if it reports a utility that is greater than or equal to the aspiration level marked by its concession strategy:
\begin{equation}
\label{eq:accept}
 ac'_{a_{i}}(X) = \left\lbrace
	  \begin{array}{l l}
	     true & \mbox{if } U_{a_{i}}(X) \geq s_{a_{i}}(t) \\
	     false & \mbox{otherwise}
	  \end{array}
	  \right.
\end{equation}

where $true$ indicates that the partial offer is acceptable at its current state for agent $a_{i}$, and $false$ indicates the opposite. A simplistic trace of a negotiation round, and how team members would behave, can be found in Fig. \ref{fig:trace}.

\section{Theoretical Analysis}
\label{sec:theo}
In this section, we analyze some of the important characteristics of our negotiation model in depth. The two main aspects that we analyze are how unanimous decisions are guaranteed regarding team decisions and how robust the proposed model is against manipulations. In the first case, the model assures that each team member gets a utility that is greater than or equal to its current aspiration level. In the second case, we analyze how the proposed model is robust against agents from the opponent, but it is easily attacked by agents from the competition that try to sabotage a deal with the opponent.
\subsection{Unanimity within the Team}
As mentioned in this article, the proposed negotiation model allows team members to reach unanimity in team's decisions. These decisions include the offer that is sent to the opponent and the acceptance/rejection of the opponent's offers. In the latter, it is clear that the proposed acceptance mechanism ensures unanimity since an opponent offer is only accepted when it is considered acceptable by all of the team members. In the former, the definition of unanimity is not straightforward.

 We define that an offer sent to the opponent $X_{A\rightarrow op}^{t}$ is a strict unanimous decision for the team when, for any team member $a_{i}$, the offer reports a utility that is greater than or equal to its current aspiration level $s_{a_{i}}(t)$:

\begin{equation}
\label{def}
 \forall_{a_{i}\in A}  U_{a_{i}}(X_{A\rightarrow op}^{t})\geq s_{a_{i}}(t) 
\end{equation}

Achieving this definition of unanimity within the team ensures that if a final agreement is found, it reports a utility that is greater than or equal to each agent's private reservation utility. In order to achieve the proposed definition of unanimous decision, some assumptions have to be made regarding the behavior of team members. These assumptions have already been presented in this article. Basically, team members have to be truthful in their responses to the mediator, following the behavior specified in Eq. \ref{eq:bid1},\ref{eq:bid2} and \ref{eq:accept}. Next, we prove that, if team members follow these behaviors, unanimity is achieved in team's decisions according to Equation \ref{def}.

\begin{IEEEproof}
 $\forall_{a_{i}\in A}  U_{a_{i}}(X_{A\rightarrow op}^{t})\geq s_{a_{i}}(t)$\\
subject to: Eq. \ref{eq:bid1}, Eq. \ref{eq:bid2}, Eq. \ref{eq:accept}, and the same type of monotonicity for valuation functions $V_{a_{i},j}$ in team members' utility functions. For the sake of simplicity, we assume that team members' valuation functions are monotonically increasing for any negotiation attribute. It should be pointed out that, in that case, the aggregation operation carried out by the trusted mediator is the $max$ operator. In any case, for any attribute $j$, its value will be determined as $x_{j}=max(x_{a_{1},j},x_{a_{2},j},...,x_{a_{M},j})$ and then it holds true that $\forall a_{i}\in A, w_{a_{i},j}V_{a_{i},j}(x_{j})\geq w_{a_{i},j}V_{a_{i},j}(x_{a_{i},j})$. The proof is quite straightforward. When the mediator declares that an attribute $j$ must be set, three different situations may arise for an agent $a_{i}$:
\begin{itemize}
 \item $a_{i}$ has already reached its aspiration level with the partial offer $U_{a_{i}}(X_{A \rightarrow op}^{'t})\geq s_{a_{i}}(t)$. Therefore, the value determined for $x_{j}$ will add utility to the partial offer and the utility reported to $a_{i}$ will further exceed its aspirations $U_{a_{i}}(X_{A\rightarrow op}^{'t})+w_{a_{i},j}V_{a_{i},j}(x_{j})\geq s_{a_{i}}(t)$. 
 \item $a_{i}$ can reach its current aspiration level $s_{a_{i}}(t)$ if it asks for a value $x_{a_{i},j}$. Thus, $U_{a_{i}}(X_{A\rightarrow op}^{'t})+w_{a_{i},j}V_{i,j}(x_{a_{i},j})=s_{a_{i}}(t)$. Since the aggregation operation is $x_{j}=max(x_{a_{1},j},x_{a_{2},j},...,x_{a_{M},j})$, the new partial offer will have a utility that is equal to or greater than its aspirations, $U_{a_{i}}(X_{A\rightarrow op}^{'t})+w_{a_{i},j}V_{a_{i},j}(x_{j})\geq s_{a_{i}}(t)$. 
 \item $a_{i}$ cannot reach its aspirations by just setting $x_{j}$. In this case, $a_{i}$ will demand the maximum possible value for $j$ and then $x_{j}=x_{a_{i},j}$. $a_{i}$ will have to reach its aspiration level by adjusting the next attributes in the agenda. In the worst case scenario, the next attribute to be set $x_{N}$ is the last one in the agenda. This means that $a_{i}$ has demanded the maximum value for the previous attributes and succeeded in getting its desired value for them. Thus, before the last attribute is set, the utility reported by the partial offer to $a_{i}$ is $\overset{N-1}{\underset{j}{\sum}} w_{a_{i},j}$. Since $\overset{N}{\underset{j}{\sum}} w_{a_{i},j}=1$ and $0\leq s_{a_{i}}(t)\leq 1$, the agent will reach its aspiration level by demanding a value for $x_{N}$ that fulfills $V_{a_{i},N}(x_{N})\geq\frac{s_{a_{i}(t)}-\overset{N-1}{\underset{j}{\sum}} w_{a_{i},j}}{w_{a_{i},N}}$, which is ensured thanks to the morphology of the valuation functions ($0\leq V_{a_{i},j}(x)\leq 1$).
\end{itemize} 
\end{IEEEproof}
One might wonder whether or not it is reasonable to think that agents are truthful in this process. However, members are not tempted to demand lesser value for attributes since the process would not ensure that the final agreement would achieve its current aspiration level. On the other hand, it is true that agents may be inclined to demand a greater value for attributes since the process ensures that the offer will be more profitable for them. Nevertheless, it should be pointed out that, generally, if more value is demanded for attributes the offer may be less profitable for the opponent and the probabilities of reaching an agreement may be greatly reduced. This issue is studied in Subsection \ref{sub:deviation}, where we analyze whether or not team members have strong incentives to deviate from the proposed behavior.

\subsection{Manipulation within the Team}

\subsubsection{Opponent}
Here, we refer to agents that infiltrate the team in order to increase the quality of the final agreement from the point of view of the opponent party. In a negotiation team setting formed by buyers, we are concerned about the fact that some seller parties may attempt to introduce agents among team members. This way, opponents may be able to maximize their own preferences by manipulating the decisions taken by the team. However, our proposed negotiation model is robust to this kind of manipulation.

Let us imagine a situation where a negotiation team wants to buy a product and a seller has been able to infiltrate agents in the team. Due to the mechanism that is employed to build the offer sent to the opponent and the mechanism employed to decide upon whether or not to accept the opponent's offer, it is not possible for opponent agents to manipulate the decisions taken within the team. Regarding the iterated offer construction process, an opponent agent would try to demand values that are close to the preferences of the opponent. In a generic electronic commerce application, an opponent agent might demand high values for the price and the dispatch date and low values for the product quality. However, the aggregation rules employed by the trusted mediator ($max$ or $min$ depending on the type of monotonic function that represents the preferences of the team members) will ensure that team preferences prevail independently of the number of infiltrated opponent agents. As for the unanimous voting process, opponent agents might try to engage team members in accepting the opponent's offer. However, this is not possible due to the fact that as long as one team member does not support the opponent's offer, it will not be accepted. Thus, it does not require further demonstration. This is the case even in situations where the group of opponent agents is larger than the number of real team members. 

For instance, let us imagine a negotiation team, formed by 5 buyers, that negotiates with a seller. Two agents ($a_{1},a_{2}$) of such team are real buyers, whereas the other three ($a_{3},a_{4},a_{5}$) are agents infiltrated by the seller. The negotiation problem is based on two negotiation attributes, price and quality, whose domains have been scaled to $[0,1]$. The valuation function used for the price in the case of the buyers is assumed to be monotonically decreasing (buyers prefer low prices to high prices), and the type of monotonic function used for the quality is assumed to be monotonically increasing (buyers prefer high quality to low quality). Thus, the mediator uses the $min$ function to aggregate the price attribute, and the $max$ function to aggregate the quality attribute. Assuming that the opponent's valuation functions are of the opposite monotonic type to those of the buyers, team members first demand the following values for the price: $x_{a_{1},price}=0.1$, $x_{a_{2},price}=0.2$, $x_{a_{3},price}=0.9$, $x_{a_{4},price}=0.85$, $x_{a_{5},price}=1$. The mediator aggregates such values and the final value for the attribute price is $x_{price}=min(0.1,0.2,0.9,0.85,1)=0.1$, which is actually preferred by the two real buyers. Thus, even if the number of opponent agents is larger than the number of real buyers, infiltrated agents from the opponent are not able to manipulate the team. The example for the quality attribute is analogous and does not require further explanation. In the end, the preferences of real buyers will prevail over opponent agents' demands and the team is not manipulated.

\subsubsection{Competitors}
Another kind of possible manipulation is the one carried out by  competitor agents. Competitors are buyer agents (in the case that the team is made up of buyer agents) that are interested in the same product as the team. Some competitors may be interested in sabotaging team deals if that assures that competitors get better deals from the opponent. This is especially true in environments where goods or services are limited (e.g., personal sellers on Ebay). Thus, competitor agents may attempt to prevent the team from reaching an agreement with the opponent.

Even though the proposed model is robust against opponent agents, robustness is not maintained when dealing with infiltrated competitor agents. In that case, the strengths shown by the model become its weaknesses. In the voting process carried out to decide upon whether or not to accept the opponent's offer, only a single agent is needed to manipulate the process and prevent the team from accepting the opponent's offers. On the other hand, competitor agents may manipulate the offer construction phase by being highly demanding. In a generic electronic commerce application, the competitor agent would demand very low values for the price, short dispatch dates and very high product quality. This way, competitor agents make offers extremely undesirable for opponent agents, preventing the team from reaching a final agreement with the opponent. Due to the aggregation operators employed by the trusted mediator, only one competitor agent is needed to manipulate the offer construction process. Thus, this model should be employed only when team members are extremely sure that no competitor agent has infiltrated the team. It would be possible to employ sophisticated mechanisms to detect these agents; however this is a topic for future research.

\section{Empirical Evaluation}
\label{sec:exp}
In this section, we explore the impact of the different parameters of our proposed model. More specifically, we study the importance of the agenda of issues imposed by the mediator on the negotiation process, the impact of the number of decision rights that are handed over during the pre-negotiation, the empirical robustness of the model against attacks (i.e., agents from the competition that try to sabotage the negotiation), and whether or not team members have incentives to deviate from the proposed strategy.
\subsection{Studying the Impact of Intra-Team Agenda}
The mediator uses an agenda to determine which attributes are set first in the iterated building process. A reasonable heuristic is to try to satisfy team members with those attributes that are less important for the opponent. Otherwise, the resultant offer may be too demanding and the negotiation process may end in failure. Thus, ideally, the agenda should order the attributes in ascendant order of importance for the opponent.

However, it is acknowledged that the situations where the opponent may reveal its full ranking of preferences are very limited or almost non-existent. Thus, it is necessary to provide the mediator with mechanisms that approximately learn the opponent's preferences. In this article, we propose a simple learning mechanism that is based on the idea that one agent may concede less in its important attributes during the first negotiation rounds. The mechanism takes into account the offers received in the first $k$ negotiation rounds and sums up the accumulated amount of concession for each negotiation attribute. Then, the mediator orders the attributes in descendant order according to the amount of concession and it becomes the agenda of attributes for the negotiation. If the number of current rounds is lower than $k$, the agenda is built based on the available information. Thus, in the first few rounds, the learning mechanism is not expected to  accurately match the opponent's preferences; however, as the  negotiation process advances, more information is available and the learning mechanism should match the opponent's preferences better. 

\begin{figure}
 \includegraphics[width=250pt]{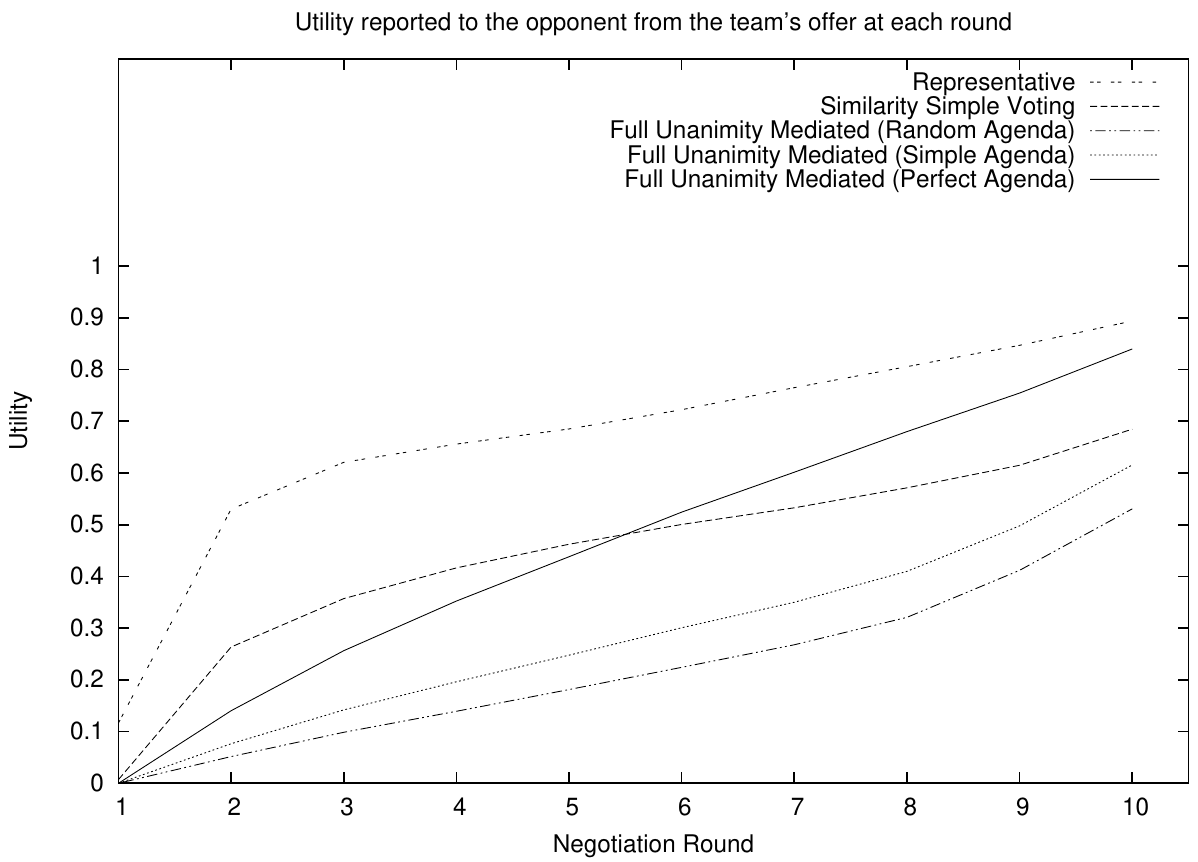}
 \includegraphics[width=250pt]{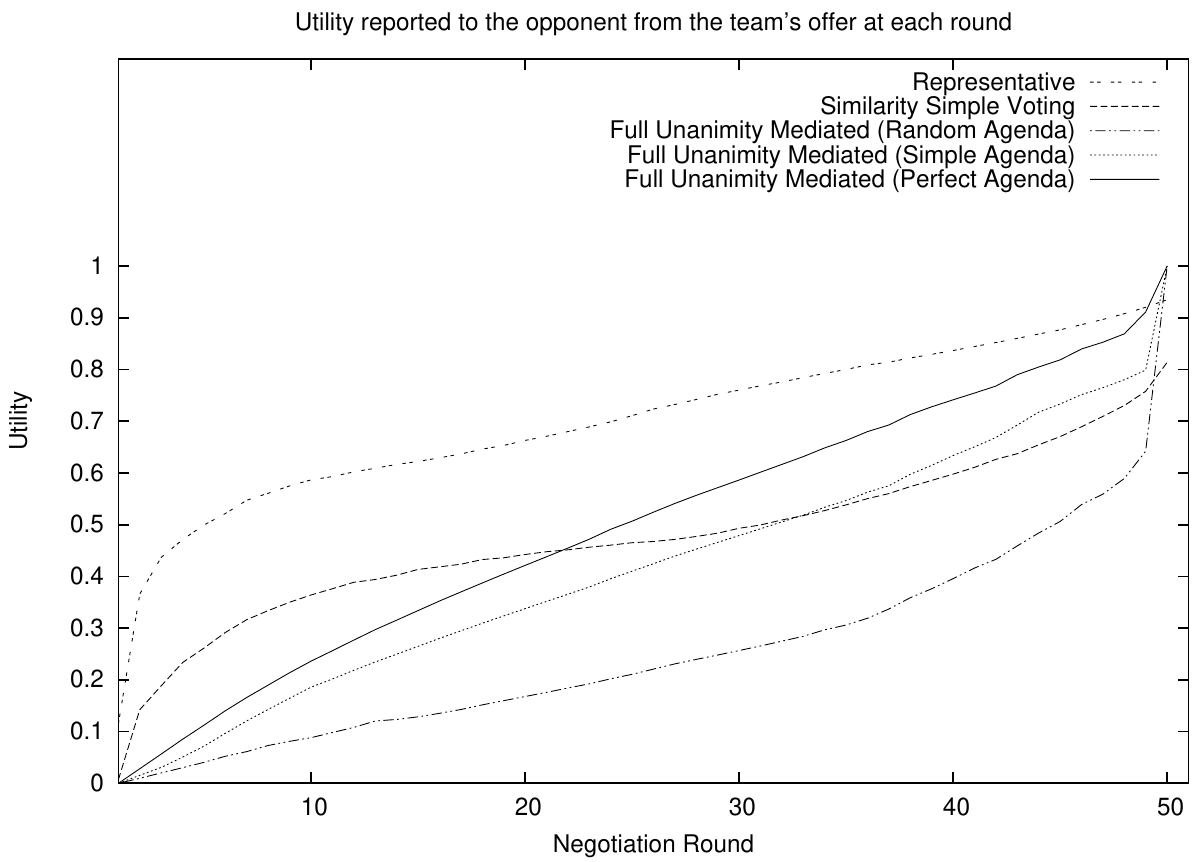}
 \caption{The upper graphic shows the average utility reported to the opponent by the team's proposal at each negotiation round for the short deadline scenario. The lower graphic shows the results for the long deadline scenario.}
 \label{fig:demanda_long}
\end{figure}

In the first experiment, we decided to study the importance of the agenda on the negotiation process. While every team member gets a utility that is greater than or equal to its desired aspiration level, the offer may be more or less demanding for the opponent. If the offer is less demanding for the opponent, it is more probable that it will be accepted by him. Therefore, we decided to study the utility reported by the teams' offer to the opponent at each negotiation round. We simulated a negotiation process where offers are not accepted (i.e., it always reaches the negotiation deadline) just to observe the utility of the offers proposed by the team from the opponent's perspective. Two different environments were tested: one with a short deadline $T_{op}=T_{A}=10$, and one with a long deadline $T_{op}=T_{A}=50$. Other parameters were set to the standard values of our negotiation model: $\beta_{op}=\beta_{A}=1$, $\epsilon_{a_{i}}=0$, and $RU_{op}=RU_{a_{i}}=0$. Three different types of agendas for the FUM model were compared: a perfect agenda where the mediator knows perfectly the order of importance given by the opponent (FUM-perfect); the simple learning method described above (FUM-simple); and a random agenda that is built at each negotiation round (FUM-random). For FUM-simple, the number of initial negotiation rounds to be taken into account was set to $k=\lfloor\frac{T_{A}}{4}\rfloor$. Additionally, the proposed negotiation model is compared with a representative model (RE) and a similarity simple voting model (SSV) which were proposed in our previous work \cite{sanchez-anguix11}. On the one hand, the representative model does not assure any kind of consensus among team members. One of the team members is chosen as representative and decides on behalf of the team according to its own private utility function. On the other hand, SSV uses majority/plurality to take team decisions. Each round, each team member is allowed to propose an offer to be sent to the opponent. This offer is proposed based on a similarity heuristic that considers the last opponent's offer and the last offer proposed by team members in the previous round. These two models are expected to be less demanding in terms of utility due to the fact that less conflict is introduced with the opponent (i.e., a fewer number of team members may reach their aspiration level).  A total of 100 random teams with size $M=4$ and random utility functions (4 attributes) were confronted with 11 randomly generated opponents. In order to capture stochastic variations in the different models, each possible negotiation was repeated 4 times. Thus, a total of 4400 negotiations were carried out per model and environment (i.e., short/long deadline). The results for this first experiment can be observed in Fig. \ref{fig:demanda_long}.

As can be observed in the short deadline scenario (Fig. \ref{fig:demanda_long}), the offers proposed by the representative model are more attractive for the opponent. This is reasonable since, in this case, the representative only negotiates attending to its own utility function. Therefore, it results in less conflict with the opponent and more trade-off possibilities. The behavior observed for the perfect agenda model and the similarity simple voting model are more surprising. Even though, in the first rounds, SSV proposes offers that report more utility for the opponent than those built by the perfect agenda model, as the negotiation advances, the perfect agenda model outperforms SSV. This happens at negotiation round 6. This may be explained by the fact that, at that point, more trade-off possibilities arise between all of the team members and the opponent, and the perfect agenda model is capable of exploiting them while assuring the desired aspiration level for each teammate. As for the simple agenda model, it performs slightly better than the random agenda model, but worse than the other methods in the experiment. This is explainable by the fact that, since the negotiation deadline is short, limited information can be used to learn the opponent's preferences. Consequently, the agenda built is closer to a random agenda than to the perfect agenda. In the case of the long deadline scenario, a similar tendency can be observed. Nevertheless, there are some differences that are worth highlighting. First, the representative model is still the one that is the most attractive for the opponent's interests. However, in this scenario, both the perfect agenda model and the simple agenda model are able to outperform SSV at some points of the negotiation process. Obviously, this happens earlier for the perfect agenda model since it represents perfect knowledge about the opponent's preferences. Hence, it is able to take advantage of possible trade-offs earlier in the negotiation. It happens approximately at round 22. Regarding the simple agenda model, it is able to outperform SSV around round 33. Differently to the first scenario, since the amount of information to learn from is greater, the simple agenda model is able to get closer to the perfect agenda and offer more attractive offers to the opponent. 

In conclusion, methods proposed in the literature like RE and SSV (in the short deadline scenario) are less demanding for the opponent; however it should be pointed out that they do not ensure unanimity as FUM. Thus, the preferences of all the team members are not represented in the deals found by RE and SSV. In fact, we ran additional tests to ascertain this conclusion. The experimental conditions were set to the same parameters found in this first batch of experiments, but in this experiment the two parties were able to accept offers, thus ending the negotiation before the deadline. We tested the performance of RE, SSV, and FUM-simple according to different quality measures such as the minimum utility of the team members and the average utility of the team members. The results of this experiment can be observed in Table \ref{tab:extra-demand}. As shown, FUM-simple is able to obtain better agreements in terms of utility (both measures) for the team members. Thus, the results suggest the aforementioned claim: even though RE and SSV are less demanding for the opponent, they do not represent the preferences of the team members as FUM does.  
\begin{table}
\centering
 \begin{tabular}{|c | c c | c c |}
\hline
                     &\multicolumn{2}{c}{Short deadline}&\multicolumn{2}{|c|}{Long deadline}\\
\hline
  Method & Min. & Ave. & Min. & Ave.  \\
\hline
  RE               & [0.10-0.11] & [0.42-0.43] & [0.11-0.12] & [0.45-0.46] \\
\hline
  SSV              & [0.33-0.34] & [0.52-0.53] & [0.30-0.31] & [0.56-0.57] \\
\hline
  FUM-simple       & [0.41-0.42] & [0.68-0.69] & [0.50-0.51] & [0.74-0.75] \\
\hline
 \end{tabular} 
\caption{This table shows the quality of the final agreement. Values shown are confidence intervals (95\%) for the mean. Min: Minimum utility of team members, Ave: Average utility of team members}
\label{tab:extra-demand}
\end{table} 
\subsection{Studying the Impact of $\epsilon_{a_{i}}$}
In this second experiment, we decided to study the impact of $\epsilon_{a_{i}}$ on the team's performance. It seems reasonable to think that low values of this parameter should help to construct offers that are more interesting for the opponent, but high values should impact negatively on the utility obtained by $a_{i}$. We devised an experiment where the value of $\epsilon_{a_{i}}$ was set in a uniform way for all of the team members. More specifically, we used the values 0, 0.02, 0.05, 0.07, 0.1, 0.12, 0.15, 0.17, 0.2 for $\epsilon_{a_{i}}$. For the quality measures, we observed the minimum and the average utility of the team members. Two different environments were tested: short/long deadline, whose lengths are drawn from the uniform distributions $T_{op}=T_{A}=U[5,10]$, $T_{op}=T_{A}=U[30,60]$, respectively. The concession speed for both parties was set to be drawn from $\beta_{op}=\beta_{A}=U[0.4,0.99]$ since initial experiments have suggested that boulware strategies may provide more utility for both parties in absence of other outside options \cite{sanchez-anguix11}. The reservation utility for the agents was drawn from a uniform distribution $RU_{op}=RU_{a_{i}}=U[0,0.25]$. In this case, the learning method for the agenda was set to FUM-simple and the number of initial rounds to be taken into account was set to $k=\lfloor\frac{T_{A}}{4}\rfloor$. A total of 100 randomly generated teams with size $M=4$ and random utility functions (4 attributes) were confronted with 12 randomly generated opponents. Each possible negotiation was repeated 4 times. Thus, a total of 4800 negotiation were carried out per model and environment. The results for this experiment are shown in Table \ref{tab:tole}.
\begin{table}
\centering
 \begin{tabular}{|c | c c | c c |}
\hline
                     &\multicolumn{2}{c}{Long deadline}&\multicolumn{2}{|c|}{Short deadline}\\
\hline
  $\epsilon_{a_{i}}$ & Min. & Ave. & Min. & Ave.  \\
\hline
  0.00               & 0.49 & 0.72 & 0.35 & 0.60  \\
\hline
  0.02               & 0.50 & 0.71 & 0.37 & 0.61  \\
\hline
  0.05               & 0.49 & 0.68 & 0.37 & 0.58  \\
\hline
  0.07               & 0.48 & 0.67 & 0.37 & 0.57  \\
\hline
  0.10               & 0.49 & 0.66 & 0.37 & 0.56  \\
\hline
  0.12               & 0.48 & 0.65 & 0.37 & 0.56  \\
\hline
  0.15               & 0.48 & 0.64 & 0.37 & 0.55  \\
\hline
  0.17               & 0.47 & 0.63 & 0.38 & 0.54  \\
\hline
  0.20               & 0.46 & 0.61 & 0.38 & 0.55  \\
\hline

 \end{tabular} 
\caption{This table shows the average impact of $\epsilon_{a_{i}}$ on team performance. Min: Minimum utility of team members, Ave: Average utility of team members}
\label{tab:tole}
\end{table} 

The results show a slight decrease in the utility (minimum utility and average utility) as $\epsilon_{a_{i}}$ gets larger. This behavior is found in almost every scenario tested. Those configurations that do not show this pattern usually obtain very similar results for all of the configurations. Thus, the agents should choose $\epsilon_{a_{i}}=0$ independently of the type of scenario where they negotiate. In the best case, the agent will get a slightly better utility than other values of the parameter.  In the worst case scenario, the agent will get a very similar utility to other values of the parameter $\epsilon_{a_{i}}$.  The value $\epsilon_{a_{i}}=0$ corresponds to the agents only handing over those decision rights associated to attributes that yield no interest at all for the agent. 

It can also be observed that the average utility is impacted more negatively by increment of the $\epsilon_{a_{i}}$ parameter in the long deadline scenario than in the short deadline scenario. A thorough analysis of our results gave an answer to this phenomenon. The results suggest that higher values of $\epsilon_{a_{i}}$ reduce the average utility for the team members. However, the number of negotiations that ended with no agreement in the long deadline scenario when $\epsilon_{a_{i}}=0$ was 151 (3.1\% of the negotiation cases ended with an average utility equal to 0), whereas the number of failed negotiations was 404 (8.41\%) when $\epsilon_{a_{i}}=0$ and the deadline was short. As $\epsilon_{a_{i}}$ was increased to 0.2, the number of failed negotiations decreased to 35 (0.7\%) in the long deadline scenario and 91 (1.8\%) in the short deadline scenario. Thus, higher values for $\epsilon_{a_{i}}$ contribute to reaching an agreement in cases where no deal was found. This effect is more notorious in the short deadline scenario. Since the number of failed negotiations is greatly reduced in the short deadline scenario, the negative effect of higher $\epsilon_{a_{i}}$ is moderated since the new negotiations contribute with values for the average utility that are greater than or equal to 0. Despite this, the reduction in the number of failed negotiations is not enough to counter the negative impact of $\epsilon_{a_{i}}$.

In general, $\epsilon_{a_{i}}$ can be considered as some sort of moderator for the initial demand. According to our results, in general, agents should not give up any decision right over an attribute that yields interest for him. Only those decision rights associated to attributes that yield no interest at all should be handed over. Hence, team members should always start demanding their highest aspiration level. This situation resembles results obtained in bilateral negotiation \cite{faratin98}, where it was found that if the deadline is reasonably long, the agent should start demanding values close to their maximum utility.
\subsection{Studying the Impact of Infiltrated Competitors}

In Section \ref{sec:theo}, we described the robustness of the proposed negotiation model against agents that try to sabotage the negotiation team. It was shown that agents from the opponent may not be able to manipulate the negotiation process. Nevertheless, when agents from the competition infiltrate the team, they may be able to stop the team from reaching an agreement. Thus, one of our concerns is how different levels of risk may affect the performance of teams acting according to our model. We decided to test the performance of FUM in different adverse negotiation scenarios. 

These scenarios differ in the probability $P$ that at least one of the team members comes from the competition. The infiltrated agent tries to stop the team from reaching an agreement with the opponent. To do so, the agent always rejects the offer received from the opponent and is extremely demanding when asking for value in the iterated offer building process. Since an agent that always asks for the most demanding value can be easily spotted (i.e., the agent does not concede at all), we decided to model this competitor agent as an agent that tries to mimic a team member with a high reservation utility $RU_{a_{i}}$. This way, the agent concedes during the negotiation process, but its requests are always high since its reservation utility is high. Hence, we decided that the infiltrated agent would follow the same concession strategy as the rest of the teammates, but the infiltrated agent would have an unexpectedly high reservation utility drawn from the uniform distribution $RU_{a_{i}}=U[0.8,1.0]$. According to this strategy, an infiltrated agent would be more difficult to identify than an agent that always asks for the most demanding value. However, this does not assure that the team will be sabotaged. This will also depend on other factors such as deadline lengths and concession strategies carried out by both parties.

In this experiment, we propose to analyze the performance of the FUM model in scenarios where $P=\{0,25,50,75,100\%\}$. Additionally, we include two modifications of FUM with two different unanimity levels: 50\% (FUM50), and 75\% (FUM75). These levels of unanimity are applied when accepting the opponent's offer. The infiltrated agent will act according to the $\beta_{A}=U[0.4,0.99]$ imposed by the team, but he will act as an agent with a high reservation utility $RU=U[0.8,1.0]$. The reservation utility for the rest of the agents was drawn from a uniform distribution $RU_{op}=RU_{a_{i}}=U[0,0.25]$. As in previous experiments, the concession speed of the opponent was set to $\beta_{op}=U[0.4,0.99]$ and we simulated two different scenarios: one with short deadlines ($T_{A}=T_{op}=U[5,10]$) and one with long deadlines ($T_{A}=T_{op}=U[30,60]$). For $\epsilon_{a_{i}}$, it was set to $\epsilon_{a_{i}}=0$ since the previous experiment showed that it may be more beneficial in terms of utility to team members.
 The learning method for the agenda was set to FUM-simple and the number of initial negotiation rounds to be taken into account was set to $k=\lfloor\frac{T_{A}}{4}\rfloor$. A total of 100 randomly generated teams with size $M=4$ and random utility functions (4 attributes) were confronted with 12 randomly generated opponents. Each negotiation was repeated 4 times. Thus, a total of 4800 negotiations were carried out per model and environment. The results are shown in Table \ref{tab:sec1}.
\begin{table}
\centering
 \begin{tabular}{|c | c | c | c | c | c | c |}
\hline
 & \multicolumn{2}{c|}{FUM}&\multicolumn{2}{c|}{FUM75}&\multicolumn{2}{c|}{FUM50}\\
\hline
 $P$                    &Long&Short&Long&Short&Long&Short\\
\hline
  0.0\%               &0.71 & 0.61 &0.67 & 0.59 & 0.58 & 0.53 \\
\hline
  25\%               & 0.60 & 0.50 & 0.65 & 0.56 & 0.58 & 0.51  \\
\hline
  50\%               & 0.47 & 0.38 & 0.61 & 0.53 & 0.56 & 0.50 \\
\hline
  75\%               & 0.33 & 0.28 & 0.59 & 0.48 & 0.55 & 0.50  \\
\hline
  100\%               & 0.24 &0.17 &0.56 & 0.46 & 0.53 & 0.48  \\
\hline
\end{tabular} 
\caption{This table shows how the probability $P$ that at least one of the team members is a competitor impacts on the team's performance. Values show the mean for the average utility of team members}
\label{tab:sec1}
\end{table} 

The results showed the expected tendency: as the probability $P$ increased, the utilitarian values for the team members decreased. This effect is observable due to the fact that the environment was more distrustful and the agents were able to successfully sabotage the team by acting as highly demanding team members. The average utility of the team members was reduced by 67\%-73\% in the highest risk scenario for FUM. As a solution for this problem, other unanimity rules could be useful. In fact, it can be observed that when $P$ is high, FUM50 and FUM75 perform better than FUM and are not so affected by the infiltrated agent (16\%-33\% performance reduction for FUM75 in the highest risk scenario, and 9\%-10\% performance reduction for FUM50).  Thus, it is acknowledged that without any additional mechanism (e.g., trust and reputation models \cite{sabater05}) the proposed negotiation model is not convenient for scenarios where it is very likely that competitor agents may enter a team. In cases where there is a high risk of encountering manipulators, models based on majority/plurality voting paradigms such as SSV \cite{sanchez-anguix11} or modifications of FUM like FUM50 and FUM75 may prove to be more fit since a large number of competitor agents may be needed to sabotage the negotiation. However, the unanimity would not be assured anymore, which we consider highly desirable for teams.

\subsection{Strategy Deviation}
\label{sub:deviation}
The proposed model assumes that team members state the truth when asked about which attribute values they need to reach their desired utility level during the offer construction phase. When dealing with selfish agents, one risk faced is the fact that selfish agents may not tell the truth in order to maximize their own utility. In this case, it seems clear that team members have no incentives to ask for less attribute value than they need since it may end up in an agreement with a utility inferior to the desired level of utility. However, team members may have incentives to demand more value if that maximizes their utilities (be more demanding). For a team member to play strategically, it would need to have some knowledge about team members' and opponent's utility functions, deadlines, reservation utilities, and other agents' strategies. We aim to propose negotiation models for open environments, where information is private. Therefore, agents usually have limited and uncertain information regarding the negotiation conditions. This leads to the question of whether or not team members would achieve higher utilities by deviating from the proposed strategy.    

In this subsection we analyze whether or not team members have incentives to deviate from the proposed strategy in the offer construction phase. For this matter, we designed two types of \textit{deviated} team members. The first type of deviated agent, which we will name \textit{slightly deviated}, behaves exactly as the standard behavior proposed for team members in this article. However, during the iterated offer construction phase, the agent does not ask for the value it needs from attribute $j$, but a value that reports higher utility than it needs. The amount of extra utility that it attempts to achieve is controlled by a parameter $d_{i}$. When $d_{i}>1$, the team member demands more value than it needs, as it can be appreciated in the formula:
\begin{equation}
\label{eq:biddeviated1}
 x_{a_{i},j}= \underset{x\in [0,1]}{\mbox{argmin }} ( d_{i}\times(s_{a_{i}}(t)-U_{a_{i}}(X_{A\rightarrow op}^{'t}))-w_{a_{i},j}V_{a_{i},j}(x))
\end{equation}
When the utility of the partial offer exceeds or equals the desired utility level $s_{a_{i}}(t)$, the agent abandons the offer construction phase at that round. The effect of this behavior is that,  when the agent is asked to set an attribute which can report the desired utility, it demands more value for that attribute and then leaves the iterated building process. For instance, if a seller agent needs 0.50 for the price attribute in order to reach its desired utility level and $d_{i}=1.25$, it will ask for $0.50\times1.25=0.625$ instead. The second type of deviated team member, named \textit{highly deviated}, behaves as the \textit{slightly deviated} team member but when it has reached its desired utility level, it stays an additional attribute in the iterated building process. When asked about the value of that extra attribute, the \textit{highly deviated} agent asks for a random value that reports between 10\% and 50\% of the attribute's utility. For instance, assuming that the price is scaled between 0 and 1, a \textit{highly deviated} seller that has reached its desired utility level would ask for a price value between 0.1 and 0.5. After setting the extra negotiation attribute, the \textit{highly deviated} team member leaves the offer construction phase.

We set the parameters of our model to the same values used in the previous experiment: $T_{A}=T_{op}=U[30,60]$ for long deadline scenarios, $T_{A}=T_{op}=U[5,10]$ for short deadline scenarios, $RU_{a_{i}}=RU_{op}=U[0,0.25]$,and $\beta_{A}=\beta_{op}=U[0.4,0.99]$. A total of 100 randomly generated teams with size $M=4$ and random utility functions (4 attributes) were confronted with 12 randomly generated opponents. Each possible negotiation was repeated 4 times. Thus, a total of 4800 negotiations were carried out per model and environment. We studied the effect of the number of \textit{slightly deviated} agents $|A|_{sd}=\{1,2,3,4\}$ (the rest of team members having the standard behavior), the effect of the number of \textit{highly deviated} agents $|A|_{hd}=\{1,2,3,4\}$ (the rest of team members having the standard behavior), and different values for $d_{i}=\{1.25,1.50,1.75\}$ (all of the deviated agents were set to have the same $d_{i}$). The quality measure studied was the average utility since an increment in the utility of one of the team members will always have a positive effect on the average utility (same type of valuation functions). The results of the experiment are depicted in Table \ref{tab:dev}. We only show the results for the long deadline scenario, but it should be noted that the same pattern was found for short deadline scenarios. It can be observed that all the combinations obtain similar results in terms of average utility. There is only a slight decrement in the average utility as we move to more demanding attitudes (e.g., $|A|_{hd}=4,d_{i}=1.75$). Even though, the differences between the most demanding behaviors and other behaviors are not large enough to be considered significant. Thus, the experimental results suggest that team members may not have incentives to deviate much from the proposed strategy. A closer look at the negotiation traces explained the previous results. While being more demanding may obtain higher utilities in successful negotiations, it may also lead to a higher number of failed negotiations, thus leading to lower or equal average utilities. These results can be observed also at Table \ref{tab:dev}, where there is a clear tendency for the number of failed negotiations to increase as team members deviate further from the standard behavior.
\begin{table}
\centering
 \begin{tabular}{c  c  c  c  c }
\hline
                     &$d_{i}=1$ & $d_{i}=1.25$ & $d_{i}=1.5$ & $d_{i}=1.75$\\
\hline
  $|A|=4$               & [0.71-0.72] & - & - & -  \\

  $|A|_{sd}=1$               & - & [0.70-0.72] & [0.71-0.72] & [0.70-0.71]  \\

  $|A|_{sd}=2$               & - & [0.71-0.72] & [0.71-0.72] & [0.70-0.72]  \\

  $|A|_{sd}=3$               & - & [0.71-0.72] & [0.70-0.72] & [0.69-0.70]  \\

  $|A|_{sd}=4$               & - & [0.70-0.72] & [0.70-0.72] & [0.69-0.71]  \\

 $|A|_{hd}=1$               & - & [0.70-0.72] & [0.71-0.72] & [0.70-0.71]  \\

 $|A|_{hd}=2$               & - & [0.70-0.71] & [0.70-0.72] & [0.69-0.71]  \\

 $|A|_{hd}=3$               & - & [0.69-0.71] & [0.69-0.71] & [0.69-0.70]  \\

 $|A|_{hd}=4$               & - & [0.69-0.70] & [0.69-0.70] & [0.68-0.69]  \\
\hline
                     &$d_{i}=1$ & $d_{i}=1.25$ & $d_{i}=1.5$ & $d_{i}=1.75$\\
\hline
  $|A|=4$               & 206 & - & - & -  \\

  $|A|_{sd}=1$               & - & 208 & 205 & 199  \\

  $|A|_{sd}=2$               & - & 202 & 230 & 236  \\

  $|A|_{sd}=3$               & - & 199 & 240 & 243  \\

  $|A|_{sd}=4$               & - & 248 & 267 & 287  \\

 $|A|_{hd}=1$               & - & 202 & 189 & 236  \\

 $|A|_{hd}=2$               & - & 241 & 246 & 274  \\

 $|A|_{hd}=3$               & - & 256 & 301 & 302  \\

 $|A|_{hd}=4$               & - & 299 & 292 & 324  \\
\hline
 \end{tabular} 
\caption{The upper table shows confidence intervals for the average utility depending on the number of \textit{deviated} agents and $d_{i}$. The lower table shows the number of failed negotiations for each case. Some of the combinations are empty since they do not make sense in practice (e.g., 0 \textit{deviated} agents and $d_{i}>1$).}
\label{tab:dev}
\end{table} 
\section{Related Work}
\label{sec:related}
In the last few years, there has been growing interest in multiagent systems as a support for complex and distributed systems. Among these complex systems, there is special interest in scenarios where multiple agents, with possibly conflicting goals, cooperate with each other to reach their own goals. The benefits of cooperation and coordination are well known, and, as stated by Klein \cite{klein98}, computer systems may help us to identify and apply the appropriate coordination mechanism. Due to the inherent conflict among agents, techniques that allow agents to solve their own conflicts and cooperate are needed. This need is what has given birth to a group of technologies which have recently been referred to as agreement technologies \cite{sierra11}. Trust and reputation \cite{sabater05}, norms \cite{dignum99}, agent organizations \cite{horling04,esparcia11}, argumentation \cite{rahwan03,pajares11} and automated negotiation \cite{jennings01,sanchez-anguix11b} are part of the core that makes up this new family of technologies.

Despite being part of agreement technologies, automated negotiation has been studied by scholars for a few years. Automated negotiation consists of an automated search process for an agreement between two or more parties where participants exchange proposals. Two different research trends can be distinguished in automated negotiation models. The first type of model aims to calculate the optimum strategy given certain information about the opponent and the negotiation environment \cite{serrano03,fatima06}. The second type of model encloses heuristics that do not calculate the optimum strategy but obtain results that aim to be as close to the optimum as possible \cite{faratin98,jonker01,faratin02,lai08}. These models assume imperfect knowledge about the opponent and the environment, and aim to be computationally tractable while obtaining good results. This present work can be classified into the latter type of models.

Most of the research has concentrated on bilateral models where each party is a single individual. Our article studies bilateral negotiations where at least one of the parties is a negotiation team, made up of more than a single individual. It should be noted that the problem of finding an agreement for a negotiation team is inherently complex since it not only requires finding an agreement with the other party but it also entails reaching some type of consensus within the team. Even though communications with the opponent party may be similar to classic bilateral models, negotiation teams may require an additional level of negotiation that involves team members. Thus, classical bilateral models cannot be applied directly if a certain level of consensus is necessary regarding team decisions. As far as we know, our previous work is  \cite{sanchez-anguix11} is the only work that focuses on negotiation teams. Despite that, bilateral negotiation is perhaps the most similar topic to our current research. Hence, we describe some of the most important negotiation models that use imperfect knowledge. A brief overview of these models can be observed in Table \ref{tab:brief}.

\begin{table}
\center
 \begin{tabular}{|c | c | c | c |}
\hline
\textbf{Work} & \textbf{Teams} & \textbf{Mediated} & \textbf{Parties} \\
\hline
Faratin et al. \cite{faratin98,faratin02} & No & No & 2\\
\hline
Jonker et al. \cite{jonker01} & No & No & 2\\
\hline
Lai et al. \cite{lai08} & No & No & 2 \\
\hline
Sanchez-Anguix et al. \cite{sanchez-anguix11b} & No & No & 2\\
\hline
Ehtamo et al. \cite{ehtamo01} & No & Yes & $n$\\
\hline
Klein et al. \cite{klein03} & No & Yes & $n$\\
\hline
Ito et al. \cite{ito10} & No & Yes & $n$\\
\hline
Sanchez-Anguix et al. (2011) \cite{sanchez-anguix11} & Yes & Yes & 2\\
\hline
 \end{tabular} 
\caption{A brief overview of state-of-the-art negotiation models and their features}
\label{tab:brief}
\end{table}  

Faratin et al. \cite{faratin98} propose a \textit{non-mediated} bilateral negotiation model for service negotiation where agents apply and mix different concession tactics (i.e., time-dependent, imitative and resource-dependent).  In their work, they analyze the impact of the model's parameters and determine which configurations work better in different scenarios by means of experiments. Our proposed work also assumes the use of time-dependent concession strategies for the calculation of agents' aspirations at each negotiation round. Additionally, we also take an experimental approach to validate the impact of our model's parameters. Later, the authors proposed a \textit{non-mediated} bilateral negotiation model \cite{faratin02} whose main novelty was the use of trade-offs to improve agreements between two parties. A trade-off consists of reducing the utility obtained from some negotiation issues with the goal of obtaining the same exact utility from other negotiation issues. The rationale behind trade-offs is to make the offer more likable for the opponent while maintaining the same level of satisfaction for the proposing agent. For that purpose, the authors propose a fuzzy similarity heuristic that proposes the most similar offer to the last offer received from the opponent. Our model does not leave room for trade-offs since the offer that is calculated at each negotiation round is deterministic with respect to the agenda of issues and the current aspiration level of team members. However, its main strength lies in the fact that it is capable of guaranteeing the desired level of utility for each team member at each round.

Jonker and Treur propose the Agent-Based Market Place (ABMP) \textit{non-mediated} model \cite{jonker01} where agents, engage in bilateral negotiations. ABMP is a negotiation model where proposed bids are concessions to previous bids. The amount of concession is regulated by the concession factor (i.e., reservation utility), the negotiation speed, the acceptable utility gap (maximal difference between the target utility and the utility of an offer that is acceptable), and the impatience factor (which governs the probability of the agent leaving the negotiation process). 

Lai et al. \cite{lai08} propose a \textit{non-mediated} bilateral negotiation model where agents are allowed to propose up to $k$ different offers at each negotiation round. Offers are proposed from the current iso-utility curve according to a similarity mechanism that selects the most similar offer to the last offer received from the opponent. The selected similarity heuristic is the Euclidean distance since it is general and does not require domain-specific knowledge and information regarding the opponent's utility function. Results showed that the strategy is capable of reaching agreements that are very close to the Pareto Frontier. Sanchez-Anguix et al. \cite{sanchez-anguix11b}  proposed an enhancement for this \textit{non-mediated} strategy in environments where  computational resources are very limited and utility functions are complex. It relies on genetic algorithms to sample offers that are interesting for the agent itself and creates new offers during the negotiation process that are interesting for both parties. Results showed that the model is capable of obtaining statistically equivalent results to similar models that had the full iso-utility curve sampled, while being computationally more tractable. 

Another topic that resembles team negotiations are multi-party negotiations. Several works have been proposed in the literature along this line \cite{ehtamo01,klein03,ito10}. For instance, Ehtamo et al. \cite{ehtamo01} propose a \textit{mediated} multi-party negotiation protocol which looks for joint gains in an iterated way. The algorithm starts from a tentative agreement and moves in a direction according to what the agents prefer regarding some offers' comparison. Results showed that the algorithm converges quickly to Pareto optimal points. Klein et al. \cite{klein03} propose a \textit{mediated} negotiation model which can be extended to multiple parties. Their main goal is to provide solutions for negotiation processes that use complex utility functions to model agents' preferences. The negotiation attributes are no longer independent, and, thus, preference spaces cannot be explored as easily as in the linear case. Later, Ito et al. \cite{ito10} proposed different types of utility functions (cube and cone constraints) and multiparty \textit{mediated} negotiation models that obtain good quality results for the proposed utility functions. The main difference between our work and multi-party negotiations lies in the nature of the conflict and how protocols are devised. Even though each team member could be viewed as a participant in a multi-party negotiation with the opponent, it is natural to think that team members' preferences are more similar (e.g., a team of buyers, a group of friends, etc.) and they trust other teammates more than the opponent (i.e., they may share more information). Furthermore, multi-party negotiation models may be unfair for agents that are alien to the team if the number of team members exceeds the number of other participants. In that case, multi-party models may be inclined to move the negotiation towards agreements that maximize the preferences of team members.

Multi-agent teamwork is also a close research topic. Agent teams have been proposed for a variety of tasks such as Robocup \cite{stone99}, rescue tasks \cite{kitano01}, and transportation tasks \cite{jennings95}. However, as far as we know, there is no published work that considers teams of agents negotiating with an opponent. Most works in agent teamwork consider fully cooperative agents that work to maximize shared goals. The team negotiation setting is different since, even though team members share a common interest related to the negotiation, there may be competition among team members to maximize one's own preferences.

Finally, given the results obtained by our proposed model, consensus building should be mentioned as a close research topic outside agent research. Our proposed model is capable of attaining consensus/unanimity regarding team decisions under the assumption of private information. Consensus building works like \cite{cook96,herrera02} usually take the assumption that all the information regarding parties is available to a trusted mediator. Cook et al. \cite{cook96}, generalize the use of distance-based measures to obtain consensus over multiple decision makers with ordinal preferences by assigning utility weights to ordinal positions. They show that this representation is equivalent to the commonly used model of using ordinal positions as utility weights for options. Herrera-Viedma et al. \cite{herrera02} propose a computational model that is able to help humans/experts to reach soft consensus over a set of alternatives. The model is based on an iterative process where two measures are used to achieve this result: a soft consensus measure, and a proximity measure. Both measures are used to evaluate how close the individual expert opinion with respect to the collective opinion is, and help the computational system to provide feedback to experts that are far from the group's opinion. Even though the goal is similar to our work, the assumptions are different. We advocate for open systems like e-commerce systems, where agents act semi-automatically on behalf of their users. Since any type of agent can be found in open environments, privacy is a big concern due to distrust and risk of exploitation. Thus, it is not possible for a mediator to know the preferences and all of the information about the participants in the negotiation. In our approach, we only consider that a limited amount of information is transmitted to the mediator.

\section{Conclusions and Future Work}
In this article, we have proposed an agent-based negotiation model for negotiation teams that interact with an opponent using the bilateral alternating protocol in electronic systems. A negotiation team is a group of two or more agents that join together as a single negotiation party because they share a common goal which is related to the negotiation process. Thus, as a team, they have to decide which offers are sent to the opponent and whether or not the offers received from the opponent are acceptable. The main strength of our proposed model lies in the fact that decisions within the team are unanimous (i.e., the utility reported by the decision is greater than or equal to the desired utility level by each team member). The negotiation model relies on a trusted mediator that coordinates voting processes, regulates an iterated process for offer construction, and guarantees unanimity.

After describing our proposal, we have shown how the model is capable of ensuring unanimity regarding team decisions. Then, we theoretically analyzed the robustness of our model against different types of attacks. The proposed model is robust against manipulations from the opponent, but it is sensitive to manipulations coming from agents from the competition that try to sabotage a possible agreement with the opponent. We also presented different experiments analyzing the impact of different parameters of the model such as the negotiation agenda followed by the mediator to decide which attributes are set first, and the impact of the number of attribute decision rights that are handed over by team members prior to the negotiation. Additionally, we carried out an empirical evaluation of the robustness of the proposed model against attacks from agents that represent the competition, which reflects our initial concerns in the theoretical analysis. Finally, we studied whether or not team members have incentives to deviate from the proposed strategy. Empirical results suggest that there is not much incentive to deviate from the proposed strategy since deviations may impact the number of failed negotiations and, thus, the average utility.

Future work includes the evaluation of the present negotiation model and other models proposed in the literature \cite{sanchez-anguix11} in different negotiation scenarios. The rationale behind this analysis is to determine which strategies are more appropriate for team members according to different criteria such as utilitarian measures (minimum, average, maximum) and computational measures (number of messages exchanged, number of negotiation rounds). Our goal is using this knowledge in a decision-making mechanism that allows teams to select the most appropriate negotiation model according to their needs.

We acknowledge that the current model is capable of reaching unanimity given the assumption of monotonic valuation functions and linear utility functions. Therefore, future work also includes exploring aggregation mechanisms that reach consensus/soft consensus for non-monotonic attributes like colors, brands, etc. In this sense, fuzzy consensus/similarity measures as the ones proposed in \cite{faratin02,herrera02} can help to aggregate agents' opinions over this type of attributes. 

Moreover, in the last few years there has been a growing interest in modeling negotiation as a dynamic process \cite{dacosta00,parachuri09}. Despite being an interesting approach, its study is still at early stages and focused on simple negotiations, while our current model involves a negotiation team, and a negotiation party, which increments the complexity of the modeling problem. In this sense, the use of new search and optimization algorithms like gravitational search algorithms \cite{askari11}, pomdp \cite{parachuri09}, and machine learning approaches with efficient data selection \cite{zhang11} can help to further improve the state of the art in negotiation. Nevertheless, dynamic modeling of agent negotiation teams is a topic that should be studied in the future.
\label{sec:conclusions}
\ifCLASSOPTIONcaptionsoff
  \newpage
\fi



%
\bibliographystyle{IEEEtran} 
\bibliography{IEEEabrv,FUM-IEEE}
%
%
\vspace{-1.0cm}
 \begin{IEEEbiography}[{\includegraphics[width=1in,height=1.25in,clip,keepaspectratio]{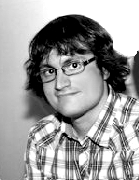}}]{Victor Sanchez-Anguix}
 was born in Valencia, Spain. He received the B.S. and M.S. degrees in Computer Science from Universidad Politécnica de Valencia, Valencia, Spain, in 2008 and 2010, respectively. He is currently a PhD. student in Computer Science at Universidad Politécnica de Valencia and holds a research grant supported by the Spanish government until 2013. His research interests include agreement technologies, soft computing, and multiagent systems.
 \end{IEEEbiography}
 \begin{IEEEbiography}[{\includegraphics[width=1in,height=1.25in,clip,keepaspectratio]{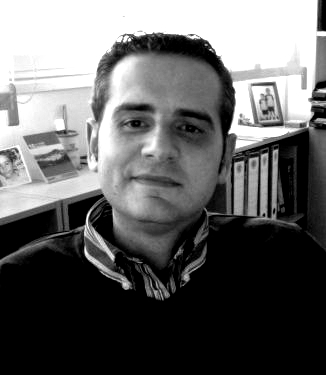}}]{Vicente Julian}
 was born in Valencia, Spain. He received his PhD. degree in Computer Science from Universidad Politécnica de Valencia in 2002. His PhD. dissertation focused on agent methodologies for real time systems. He is a researcher with \textit{Grupo Tecnología Informática - Inteligencia Artificial} at Universidad Politécnica de Valencia. He also holds an Associate Professor position. His research interests are adaptive systems, agreement technologies, and software engineering.
\end{IEEEbiography}
 \begin{IEEEbiography}[{\includegraphics[width=1in,height=1.25in,clip,keepaspectratio]{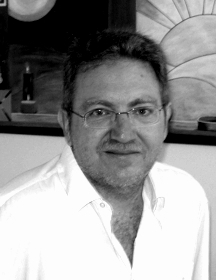}}]{Vicente Botti}
 was born in Valencia, Spain. He received his PhD. degree in Computer Science from Universidad Politécnica de Valencia in 1990. His PhD. dissertation focused on temporal knowledge based systems. He is currently head researcher with \textit{Grupo Tecnología Informática - Inteligencia Artificial} at Universidad Politécnica de Valencia. He also holds a Full Professor position. Some of his research interests include agreement technologies, and real time systems.
 \end{IEEEbiography}
 \begin{IEEEbiography}[{\includegraphics[width=1in,height=1.25in,clip,keepaspectratio]{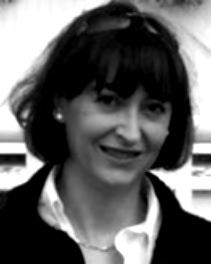}}]{Ana Garcia-Fornes}
  was born in Valencia, Spain. She received her PhD. degree in Computer Science from Universidad Politécnica de Valencia in 1996. Her PhD. dissertation focused on real time architectures for intelligent systems. She is head researcher with \textit{Grupo Tecnología Informática - Inteligencia Artificial} at Universidad Politécnica de Valencia. She also holds an Associate Professor position. Real time systems, agent infrastructures and agreement technologies are some of her interests.
 \end{IEEEbiography}



\end{document}